\documentclass[twocolumn,showpacs,preprintnumbers,amsmath,amssymb,aps,prl]{revtex4}
\usepackage{graphicx}
\begin{document}

\title{Rectification of Swimming Bacteria and Self Driven Particle Systems by 
Arrays of Asymmetric Barriers} 
\author{M.B. Wan$^{1,2}$, 
C.J. Olson Reichhardt$^{1}$, 
Z. Nussinov$^{1,2}$, 
and C. Reichhardt$^{1}$}  
\affiliation{
{$^1$}Theoretical Division,
Los Alamos National Laboratory, Los Alamos, New Mexico 87545\\
{$^2$}Department of Physics, Washington University, St Louis, 
Missouri 63160}

\date{\today}
\begin{abstract}
We show that the recent experimental observation of the rectification of 
swimming bacteria in a system with an array of asymmetric barriers
occurs due to the ballistic component of the bacteria trajectories
introduced by the bacterial ``motor.''
Each bacterium selects a random direction for motion and then moves in
this direction for a fixed period of time before randomly changing
its orientation and moving in a new direction.
In the limit where the bacteria undergo only Brownian motion
on the size scale of the barriers,
rectification 
does not occur.     
We 
examine the effects of steric interactions between the bacteria 
and observe a clogging effect upon increasing the bacteria density.
\end{abstract}
\maketitle

\vskip2pc
Recent experiments have shown that 
swimming bacteria undergo a rectification effect when placed
in a container that has an array of 
funnel-shaped barriers \cite{Galajda}. 
The initial bacteria density $\rho_b$ is constant throughout the container,
but over time, bacteria accumulate on one side of the container due to
interactions with the barriers.
The ratio 
$r=\rho^{(1)}/\rho^{(2)}$ of the bacteria density 
$\rho^{(1)}$ on one side of the container to the 
density $\rho^{(2)}$ on the other side of the container
increases over time in the presence of a single line of barriers, 
reaching a saturation value of around 
$r=2.75$.  
By taking advantage of this rectification effect, it has been
experimentally demonstrated that
various types of bacteria patterns can be induced
to form, as well as closed circuits such as a bacteria pump created
from a series of barrier arrays.
If dead bacteria or genetically modified nonswimming bacteria are placed
in the system instead of normal swimming bacteria, the rectification
effect vanishes.
The dead or nonswimming bacteria still undergo thermal motion and can be 
regarded as Brownian particles. 
In the experiments, a mixture of swimming and nonswimming bacteria was 
separated using the rectification effect of the barrier array.

The basic questions raised by these experiments are:
$(i.)$ what are the crucial 
properties necessary to give rise to rectification of the swimming bacteria
and $(ii.)$ what prevents the nonswimming bacteria from undergoing
rectification.
Possible factors that could be important include
hydrodynamic interactions with the barrier walls, the 
details of the placement of 
the driving motor on the bacteria, the elongated shape
of the bacteria, the distance that swimming bacteria move before tumbling and
selecting a new swimming direction, 
and the 
dimensions of the asymmetric barriers. Additionally, 
other complications such as 
bacterium-bacterium interactions and chemical sensing could also play a role
in the rectification.     
In Ref.~\cite{Galajda}, the rectification is attributed to a funnel mechanism
arising from interactions between the bacteria and the barriers.

Motivated by ideas in Ref.~\cite{Galajda},
in this work we study a simple model 
for pointlike swimming bacteria which produces rectification in the presence of
asymmetric barriers.
Each bacterium moves ballistically 
under the influence of a motor force in a randomly chosen direction for a 
fixed distance $l_b$ before selecting a new random direction  and
moving ballistically again.
This type of 
motion is known to occur for swimming bacteria which
move in a fixed direction for a period of time  
before undergoing a tumbling locomotion and moving in a 
new direction \cite{Berg}. 
We assume that upon interacting with a barrier, the bacterium does
not reorient but moves 
along the barrier wall at a velocity determined by the component of the
motor force that is parallel to the wall.
As a result, the bacteria are entrained along the barrier walls 
as in Ref.~\cite{Galajda}
and rectification of the bacteria density occurs. 
The rectification persists
even when an additional Langevin noise term is added 
to the ballistic motion of the bacteria.

For a system with a single line of barriers, we show that the 
magnitude of the rectification 
increases with increasing ballistic flight distance $l_{b}$,
while in the limit where the bacteria move like Brownian particles 
on the length scale of the barriers,
the rectification  vanishes. 
We also find that in this model the bacteria density is highest along the
container walls, in agreement with experimental observations.
When steric interactions between the bacteria are included, the 
rectification is reduced due to a clogging effect.
We show that the experimental system is a realization of 
a correlation ratchet \cite{Doering}, in which particles with 
fluctuations that violate detailed balance 
can give rise to a ratchet effect in the presence of an
additional asymmetry such as an asymmetric substrate. 
Our model should also be useful for other systems of self-propelled particles 
where the particle motion is non-Brownian on a certain length scale.  
Recent examples of such systems include self-propelled colloids \cite{Howse}, 
artificial swimmers made from magnetic beads \cite{Stone},
autonomously moving catalytic nanorods \cite{Mallouk}, 
and self-propelled nematics \cite{Thakur}.   
Since the amount of rectification is affected by 
the ballistic flight length $l_b$, the asymmetric barriers
could be used to sort different species
of swimming or self-propelled particles. 

\begin{figure}
\includegraphics[width=3.5in]{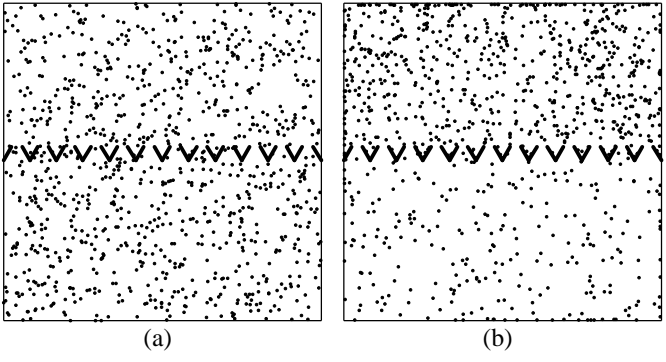}
\caption{
Images of the simulation system with $\theta=60^\circ$.  
Black dots: bacteria positions; heavy black
lines: barrier locations.
(a) Initial system configuration with uniform bacteria density for
$N_{b} = 980$ and $l_{b}=40$.
The density in the upper chamber is denoted $\rho^{(1)}$ and the 
density in the lower chamber is $\rho^{(2)}.$ 
(b) The same system after  
$100\tau$ simulation steps showing the higher bacteria density in 
the upper chamber, $\rho^{(1)}>\rho^{(2)}.$   
}
\end{figure}

Our simulation consists of a bounded two-dimensional box of 
an array of  
funnel shaped barriers as shown in Fig.~1.
The system has dimensions of $L\times L$, where $L=99$
unless otherwise noted,  and 
contains $N_{b}$ particles representing bacteria
at an overall density $\rho_b=N_b/L^2$.
The particles move in an overdamped media where we do not take into account 
any hydrodynamic effects. 
A single bacterium $i$ obeys the following equation of motion:  
\begin{equation} 
\eta \frac{d{\bf R}_{i}}{dt} = {\bf F}^{m}_{i}(t) + {\bf F}^{T}_{i} + 
{\bf F}^{B}_{i} + {\bf F}_{i}^S  
\end{equation}
Here ${\bf R}_{i}$ is the position of bacterium $i$
and $\eta$ is the phenomenological damping constant
which we set to $\eta=1$.   
The driving force from the bacterial motor 
${\bf F}^{m}_{i}(t)$
has a constant magnitude $|{\bf F}^m|=2.0$ but is applied in a randomly chosen
direction which changes after every $\tau$ simulation
time steps.  In the absence of
other forces the bacterium would move ballistically over a 
distance $l_b=\tau\delta t |{\bf F}^m|$ before reorienting,
where $\delta t=0.005$ is the magnitude of a simulation time step. 
The resulting random walk
motion is illustrated in Fig.~2(a) for $\tau=1000$ and $l_b=10$.
As $\tau\rightarrow 1$, the motion approaches
a random walk even at the smallest size scales, 
as shown in Fig.~2(b) for $l_b=1$. 
The thermal force arising from fluctuations in the solvent
is given by ${\bf F}_i^{T}$, which has the properties
$\langle F_i^{T}(t)\rangle = 0$ 
and $\langle F_i^{T}(t)F_j^{T}(t^{\prime})\rangle 
= 2\eta k_{B}T\delta_{ij}\delta(t - t^{\prime})$.  
The bacteria motion under the influence of both thermal fluctuations 
and the ballistic motor force is illustrated in Fig.~2(c)
for $l_b=10$ and $F^T=10$.
For most of this work we set $F^T=0$.
The force from the barriers 
and container walls      
is given by ${\bf F}_i^{B}$. 
Each barrier is modeled by two half-parabolic domes of strength $f_B=30$ and
radius $r_B=0.05$ 
separated by an elongated region of length $L_B=5.0$ which repels
the bacteria perpendicular to the trap axis:
${\bf F}^{B}_{i}=\sum_{k=0}^{N_B}
[\frac{f_Br_1}{r_B} \Theta(r_1){\bf \hat{R}}_{ik}^\pm
+\frac{f_Br_2}{r_B} \Theta(r_2){\bf \hat{R}}_{ik}^\perp].$
Here the total number of barriers (including confining walls) is $N_B=28$,
$r_1=r_B-R_{ik}^{\pm}$,
$r_2=r_B-R_{ik}^{\perp}$,
$R_{ik}^\pm=|{\bf R}_i-{\bf R}_k^B \pm L_B{\bf {\hat p}}^k_\parallel|$,
${\bf {\hat R}}_{ik}^\pm=({\bf R}_i-{\bf R}_k^B \pm L_B{\bf {\hat p}}^k_\parallel)/R_{ik}^{\pm}$,
$R_{ik}^{\perp}=|({\bf R}_i - {\bf R}_k^B) \cdot {\bf {\hat p}}^k_{\perp}|$,
${\bf {\hat R}}_{ik}^{\perp}=(({\bf R}_i - {\bf R}_k^B) \cdot {\bf {\hat p}}^k_{\perp})/R_{ik}^{\perp}$,
${\bf R}_i$  (${\bf R}_k^B$)
is the position of bacterium $i$ (barrier $k$),
and ${\bf \hat{p}}^k_\parallel$ (${\bf \hat{p}}^k_\perp$) is a unit
vector parallel (perpendicular) to the axis of barrier $k$.
Each funnel is modeled as two barriers meeting at a common endpoint and
placed at angles $\theta$ and $\pi-\theta$
with the $x$-axis.  Our system contains 12
funnels in a one-dimensional array with a lattice constant of $l_S=8.25$.
The steric force from the bacterium-bacterium interaction, ${\bf F}_{i}^S$, 
is modeled as a stiff repulsive spring of range 
$r_{s}=0.35$ and strength $f_s=150$, with
${\bf F}_i^S=\sum_{j\ne i}^{N_b}\frac{f_sr_3}{r_s}\Theta(r_3){\bf {\hat R}}_{ij}$.
Here $r_3=r_s-R_{ij}$,
$R_{ij}=|{\bf R}_i-{\bf R}_j|$, and
${\bf {\hat R}}_{ij}=({\bf R}_i-{\bf R}_j)/R_{ij}$.
For most of this work we consider the 
limit where bacterium-bacterium interactions are irrelevant
and set $f_s=0$.  

\begin{figure}
\includegraphics[width=3.5in]{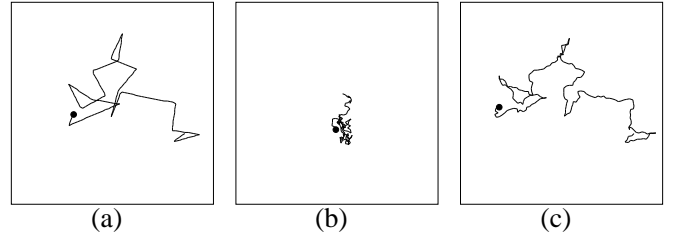}
\caption{
Dots: position of one bacterium in the system.  Lines: trajectory
of the bacterium over a fixed time period.
(a) A system with only a motor force showing ballistic motion
and a random reorientation each time the bacterium has moved a
distance $l_{b}=10$. Here $F^T=0$.  
(b) A system with small $l_{b}=1$ 
and $F^T=0$.
(c) A system with both ballistic motion and thermal motion, where
$l_b=10$ and $F^T=10$.
}
\end{figure}

We base our system geometry on the work in Ref.~\cite{Galajda}.
In Fig.~1(a) we illustrate the 
system geometry, showing the 
array of funnel shapes,
the boundary walls, and the initial
locations of the bacteria.  The bacteria density in the upper 
(lower) chamber is
$\rho^{(1)}$ 
($\rho^{(2)}$).  
The angle $\theta$ 
of each funnel barrier is set to $\theta=60^\circ$.
We first investigate the simplest set of parameters that produce 
rectification by fixing $F^T=0$, $f_s=0$, and $l_b=40.$  
The initial density ratio $r=\rho^{(1)}/\rho^{(2)}=1$, as shown in
Fig.~1(a).  In Fig.~1(b) we illustrate the same 
system after $100\tau$ time steps when $r=2.3$,
showing 
a density increase in the upper half of the sample.
A buildup of bacteria occurs along 
the barrier walls and the boundaries, 
similar to that observed in the experiments of Ref.~\cite{Galajda}.

\begin{figure}
\includegraphics[width=3.5in]{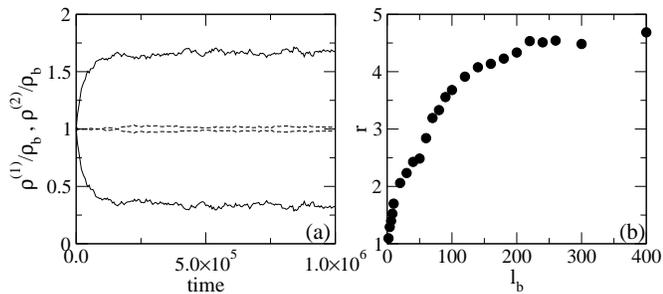}
\caption{
(a) The time dependence of the bacteria density 
$\rho^{(1)}/\rho_b$ (upper line) and $\rho^{(2)}/\rho_b$ 
(lower line), normalized by the overall bacteria density $\rho_b$, 
for the same
system in Fig.~1 with $l_b=180$, showing the density buildup over time 
in the top half of the sample.
Dotted lines: $\rho^{(1)}/\rho_b$ and $\rho^{(2)}/\rho_b$ 
for the same system but with
$l_b=0.01$ so that the bacteria undergo Brownian motion 
on the size scale of the barrier
and no rectification occurs.  
(b) $r=\rho^{(1)}/\rho^{(2)}$ versus 
$l_b$ for a sample with $\theta=60^{\circ}$
measured after $10^3\tau$ simulation steps.  
}
\end{figure}

In Fig.~3(a) we plot the time dependent behavior of 
$\rho^{(1)}/\rho_b$ and $\rho^{(2)}/\rho_b$ for the same system 
shown in Fig.~1 with $l_b=180$.  We also show the same
quantities for a system with a much smaller 
$l_{b}=0.01$ 
in which the bacteria motion is equivalent to a random walk
on the size scale of the barriers. 
For the small 
$l_b$ there is no rectification of the
bacteria density as would be expected, 
while for the large 
$l_b$ the system rectifies. 
The time dependence 
of $r=\rho^{(1)}/\rho^{(2)}$ at short times can be fit to an 
exponential form, similar to the experiments of 
Ref.~\cite{Galajda}.
We note that asymptotically, 
$r\approx 4.3$, 
while the asymptotic value 
of $r$ depends on 
$l_b$. We
have repeated the simulations
for different values of $l_{b}$, and 
in Fig.~3(b) we plot $r$ versus $l_b$ measured after $10^3 \tau$ 
simulation time steps.
For low 
$l_{b}<200$, the rectification 
increases monotonically with $l_b$, 
with $r$ reaching a maximum of value of $r=4.8$
at $l_b=200$. We note that $r$ depends in a similar way upon the
barrier length $L_B$ for fixed $l_b$,
and the rectification vanishes
for large values of $L_B$ when there is no longer a gap between 
adjacent funnel barriers. 

\begin{figure}
\includegraphics[width=3.5in]{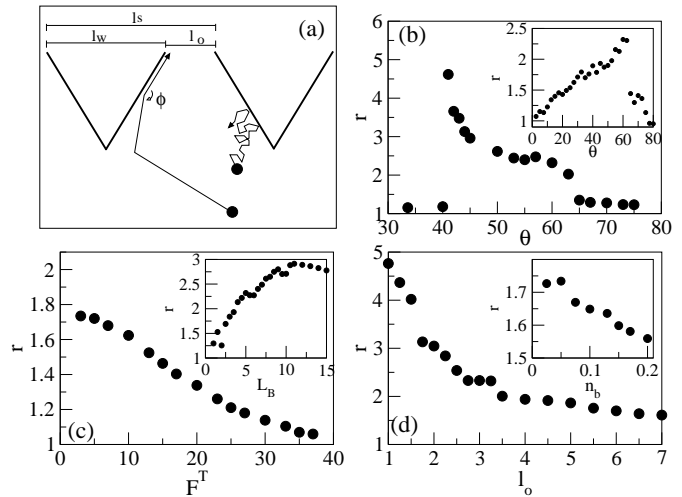}
\caption{
(a) Schematic trajectories for: (dark line) a bacterium moving 
under motor forces only and being entrained by the wall;
(light line) a bacterium experiencing only Brownian forces.
The 
barrier spacing $l_S$, opening 
size $l_o$, 
funnel width $l_w$, 
and angle $\phi$ between the bacterium trajectory and the funnel barrier wall
are also indicated.
(b) $r=\rho^{(1)}/\rho^{(2)}$ vs $\theta$ for a system with $l_{b} = 20$,
$F^T=0$,
$f_s=0$, and $L=99$.  Inset: $r$ vs $\theta$ for the same system with fixed
$l_o=3.25$ and varied $L$.
(c) $r$ vs $F^{T}$ for a system with $l_{b} = 20$, $\theta=60^{\circ}$, $L=99$, 
and $f_s=0$.  Inset: $r$ vs $L_B$ for a system with $l_b=20$, $F^T=0$,
$f_s=0$, fixed $l_o=3.25$, and varied $L$.
(d) $r$ vs $l_o$ for a system with $l_b=20$, $\theta=60^{\circ}$, $L_B=5.0$,
$f_s=0$, and varied $L$.
Inset: $r$ vs overall bacteria
density $\rho_{b}$ for a system with steric interactions,
$f_s=150$, $\theta=60^{\circ}$, $f^T=0$, $L=99$, and $l_{b} = 20$.  
}
\end{figure}

These results suggest that hydrodynamic effects or chemotaxis are not 
required to produce the rectification seen in Ref.~\cite{Galajda},
but that instead the rectification arises from the fact 
that the bacteria move in finite steplike motions.  
In Fig.~4(a) we illustrate the specific mechanics that give 
rise to the funneling behavior 
of the bacteria at the barrier. 
As described in Ref.~\cite{Galajda},
a bacterium which collides with a barrier cannot continue to move 
in the direction of the motor force; instead, the bacterium slides along
the barrier at a velocity determined by the component of the
motor force 
parallel to the barrier. 
If 
$l_b$ is large enough, there is a high probability 
that the bacterium will slide all the way across the barrier to the
other half of the container before the motor force changes its orientation.
If 
$l_b$ is very small, the bacterium cannot move very far along the
barrier before it randomly moves off when the motor force changes,
as illustrated by the light line in Fig.~4(a).  The 
rectification is produced by the combination of the asymmetric barrier and
the motion of the bacteria that allows a breaking of detailed balance 
when the bacteria interact with the barriers.
Another consequence of this model is that the bacteria should 
tend to accumulate on the container barrier walls, as 
observed in the experiment 
\cite{Galajda}.  
We note that the real bacteria 
may tend to be hydrodynamically trapped by the wall.
This effect would tend to enhance the rectification effect 
since the bacteria could move 
for distances longer than $l_b$ along the barrier,
generating rectification even for 
moderately small $l_b$.

To compare our simulation results quantitatively to experiment, we
take a simulation distance unit of 1 micron and time step unit of
$5 \times 10^{-4}$ sec.
This gives a bacteria velocity of 20 microns/sec, close to the estimated speed
$v \approx 23$ microns/sec for {\it E. coli} \cite{Mittal,Darnton}.
A tumbling rate of roughly 1 Hz \cite{Mittal} corresponds to
$l_b=20$ microns; Ref.~\cite{Galajda} gives the estimate $l_b \approx 50$ 
microns.
Fig.~3(a) shows that our system reaches a stationary 
state within $1 \times 10^5$ time steps,
which corresponds to 50 minutes.  This compares well with the saturation
time scale in the experiment of 20 to 80 minutes \cite{Galajda}.

We next consider changing the geometry of the barrier funnels.
In Fig.~4(b) we plot $r$ vs $\theta$, 
the funnel angle,
after 
$10^3$ simulation steps for 
a system with $l_{b} = 20$ and fixed $L=99$. For $\theta < 41^{\circ}$ 
adjacent funnel barriers are in contact
so there is no transport.  The maximum of $r=4.6$ occurs at 
$\theta=41^{\circ}$, and $r$ decreases with increasing $\theta$ for 
$\theta>41^{\circ}$, reaching $r=1$ at $\theta=90^{\circ}$.
The experiments also showed no rectification for flat barriers
\cite{Galajda}. 
The inset of Fig.~4(b) shows the dependence of $r$ on $\theta$ when the
space between funnel barriers $l_o$ is held fixed, indicating that the
maximum rectification occurs for $\theta=60^\circ$.  The inset of Fig.~4(c)
illustrates that $r$ initially increases with barrier length $L_B$ before
saturating, while Fig.~4(d) shows that $r$ decreases as $l_o$ increases.

A simple heuristic expression gives 
a lower bound for the value of $r$ in the case
when the spacing $l_S$ between adjacent funnel barriers 
is much smaller than the ballistic
flight length, $l_S \ll l_b$.
Assuming that the bacteria are entrained by any barrier wall which
they strike, detailed balance implies that 
in the steady state at long times,
\begin{eqnarray}
r=\rho^{(1)}/\rho^{(2)} =
Prob.(\phi > \pi/2)(l_w/l_o) + 1.
\label{detail_balance}
\end{eqnarray}
Here, 
the tip-to-tip width 
of a single funnel barrier is $l_w=l_S-l_o$, as indicated in Fig.~4(a).
In Eq.(\ref{detail_balance}), 
$Prob.(\phi > \pi/2)$ is the probability that whenever the 
bacteria hit the funnel barrier walls, they do so at a large enough angle 
$\phi$ such that they slide along the wall in an upward direction
to emerge in the top portion of the sample, as shown in Fig.~4(a).
Equation (\ref{detail_balance}) 
leads to a value of $r$ which is strictly bounded
from below by one, reflecting the rectification effect. 
By measuring $Prob.(\phi > \pi/2)$ 
we have verified that Eq.(\ref{detail_balance}) is indeed 
generally satisfied. 

In the experimental system, the bacteria experience some additional 
Brownian fluctuations due to the solvent.
We model this effect by setting $F^T>0$, producing the type of motion
shown in Fig.~1(c).
If only the thermal noise term is applied and the motor force 
$F^{M} = 0$, we find no rectification.
In Fig.~4(c) we show $r$ vs $F^{T}$ for a system with $l_{b} = 20$
and barrier radius $r_b=0.5$. 
As $F^{T}$ increases, the rectification is slowly reduced, 
while for large enough $F^{T}$ the rectification is washed out and $r=1$.
The destruction of the rectification occurs because the thermal term 
causes some of the bacteria to move away from the barriers rather than
channeling along them, so the progress along the walls is reduced 
from the mechanism shown in Fig.~4(a).
 
The bacteria in the experimental system have a steric interaction 
with each other which prevents overlap. 
To mimic this effect we add a steric force interaction by setting $f_s>0$.
In Fig.~4(d) we plot $r$ vs bacteria density $\rho_{b}=N_b/L^2$ 
for the same system in Fig.~2
but with a finite steric interaction strength of $f_s=150$. 
As the density $\rho_b$ increases, the effective rectification is 
reduced due to a clogging effect since only a small number
of bacteria can fit
along the barrier wall. 
The effect of the clogging can be reduced by the addition of thermal 
noise since this has a tendency to prevent bacteria from clustering.   
 
We have also performed simulations for mixtures of interacting bacteria 
in which one species moves under both a motor force and thermal noise
while the other species moves only due to thermal noise.
This mimics a system with a combination of swimming and nonswimming 
bacteria. If the density of the nonswimming bacteria is sufficiently high, the
rectification efficiency is reduced. It is also possible for an entrainment 
effect to occur in which the swimming bacteria can induce a weak 
rectification of the nonswimming     
bacteria.

The rectified motion in this system has similarities to   
Brownian ratchets where particles undergoing Brownian motion
in the presence of an asymmetric potential 
can exhibit a net drift  in the presence
of an additional ac drive or flashing potential substrate   
\cite{Reimann}. Our system is not a Brownian ratchet
in this sense; instead, it can be considered to be 
a realization of a correlation
ratchet. In correlation ratchets, an overdamped particle
can exhibit dc drift on an asymmetric substrate in the absence of an
ac flashing or rocking provided that the fluctuations of the
particle motion
have certain properties that break detailed balance \cite{Doering}.
In bacteria and other self-propelled particle systems, 
it is the motor force that causes the fluctuations to break
detailed balance.    

To summarize, we show that a rectification phenomenon similar to that 
observed in recent experiments \cite{Galajda} can be achieved in 
a simple model for overdamped swimming bacteria 
where the bacteria move in straight lines 
of length $l_{b}$ before  
randomly changing direction while also interacting with funnel shaped 
barriers.
In this model, no hydrodynamic interactions are needed to produce 
the rectification.  In the limit of bacteria 
which undergo only Brownian motion on the size scale
of the barriers, no rectification occurs,
in agreement with experiment \cite{Galajda}. 
The rectification results when $l_{b}$ is large enough 
that bacteria which strike a funnel barrier move along 
the entire length of the barrier before the bacterial 
motor force changes direction.
We have also examined the effects of 
steric interactions on the rectification and find a clogging 
phenomenon at the barrier funnels which reduces
the effectiveness of the rectification. 
Additionally, if a sufficiently strong Langevin noise term
is added to the model, the rectification is reduced.
This system can be regarded as a two-dimensional realization of a 
correlation ratchet in which
fluctuations that break detailed balance can give rise to directed
motion in the presence of an asymmetric substrate.   
Our results should also be relevant to other 
two-dimensional systems of self-propelled
particles that undergo a similar type of motion.  

This work was carried out under the auspices of the 
NNSA of the 
U.S. DoE
at 
LANL
under Contract No.
DE-AC52-06NA25396.

\end{document}